# PRINCIPLE OF VIRTUAL USE METHOD IN COMMON GATEWAY INTERFACE PROGRAM ON THE DACS SCHEME

Kazuya Odagiri[1], Shogo Shimizu[2], Naohiro Ishii[3], MakotoTakizawa[4]

[1] Yamaguchi University,Ymaaguchi-shi, Ymaguchi, Japan
`odagiri@yamaguchi-u.ac.jp, kazuodagiri@yahoo.co.jp`
[2] Advanced Institute of Industrial Technology, Tokyo, Japan
[3] Aichi Institute of Technology, Aichi, Japan
[4] Seikei University,Tokyo, Japan

*ABSTRACT*

*In the world of the Internet, Web Servers such as Apache and Internet Information Server (IIS) were developed to exchange information among client computers having different Operation System. They have only the function of displaying static information such as HTML files and image files into the Web Browser. However, when the information is updated, the administrator updates it by manual operation. In some cases, because it is necessary to update several places about the same information, the work load becomes high than it is assume and update error and update omission may occur. These problems were solved by use of a Common Gateway Interface (CGI) program such as a bulletin board system and a Blog system. However, these programs opened to Internet have often no user authentication mechanism and no access control mechanism. That is, they have the problem that user can access it freely only by getting the URL and inputting it to a Web Browser. Therefore, in this paper, we show a method to add the user authentication and access control mechanism for them. It is called virtual use method of CGI and is realized in the case of introducing the Destination Addressing Control System (DACS) Scheme, which is a kind of Policy Based Network Management Scheme (PBNM). As the result, this kind of the CGI program can be used in the organization with the above two functions..*

## 1. INTRODUCTION

In the world of the Internet, Web Servers such as Apache and Internet Information Server (IIS) are developed to exchange information among client computers having a different Operation System. However, these Web Servers have only the function of displaying static information such as HTML files and image files into the Web Browser. In this case, same information is often described at the several places in them. When the information needs to be updated, all duplicate information must be updated by an administrator accurately. But, the work load becomes high than it is assume and update error and update omission may occur. To cope with this problem, the Common Gateway Interface (CGI) [25] was produced.

By using the CGI, update error and update omission were prevented. These problems were solved by use of a Common Gateway Interface (CGI) program such as bulletin board system and access counter, Blog system or Wiki. However, these programs opened to Internet have often no user authentication mechanism and no access control mechanism. That is, they have the problem that

DOI : 10.5121/ijcnc.2012.4111     147

user can access it freely only by getting the URL and inputting it to a Web Browser. We define the above CGI program as Free Use CGI Program (FUCGIP).

In this paper, we show a method to add the user authentication and access control mechanism for it, which is called virtual use method of CGI. By using this method, different users become possible to use a FUCGIP virtually by inputting same URL into a Web Browser. For example, to let only authorized user group use a bulletin board system as the FUCGIP, the user authentication and access control mechanism must be added by customization. However, in this method, after different name directories are created from the original directory with programs of the bulletin board system by cut and paste and the related setting change are performed, different users can access the programs in the above different directories by inputting same URL into the Web Browser with the effect that is equal to the user certification and access control.

This method is realized in the case of introducing the Destination Addressing Control System (DACS) Scheme as a network management scheme. The DACS Scheme is a kind of Policy Based Network Management Scheme (PBNM) that we have been proposed before. This scheme is used to manage a local area network (LAN) through the communication controls on client computers. As the works of DACS Scheme, we showed the basic principle of the DACS Scheme [22], and a security function [23]. After that, we implemented a DACS System to realize a concept of the DACS Scheme [24].

The rest of paper is organized as follows. Section 2 shows related technologies and motivation of this research. In Section 3, we describe the content of the DACS scheme. Then, in Section 4, the content of DACS Web Service is explained. In Section 5, virtual use of the CGI program is shown.

## 2. RELATED TECHNOLOGIES AND MOTIVATION

The CGI program as FUCGIP has the mechanism which stores the information in a database and creates the Web page dynamically to show information necessary for each user. The scheme of the CGI is standardized as RFC3875 [26] and defined by Internet Engineering Taskforce (IETF). The basic mechanism of the scheme is described by the following figure 1.

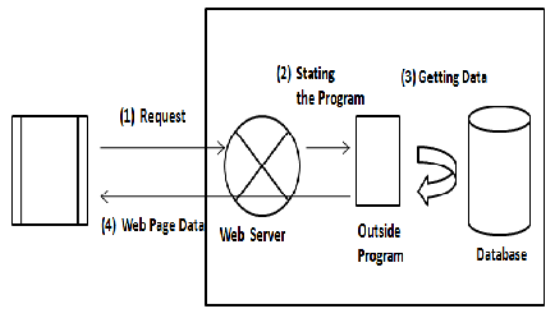

Figure 1 CGI Scheme

Because virtual usage method of FUCGIP is shown in this paper, explanations about virtualization technology are performed. First, as one of the virtualization technologies, server virtualization is listed. There are two methods in server virtualization.

(1)     Server OS virtualization

(2)     Server hardware virtualization

Server OS virtualization is the technology that one server OS called host OS divides hardware resources such as CPU, RAM into multiple virtual OS's called container. This is a technology to show so that the multiple OS's operate in a server machine. As software to realize it, OpenVZ [27], Virtuozzo [28], Linux Container [29] and Solaris Container [30] are listed.  As merits, compared with server hardware virtualization explained later, the points that consumption of the hardware resource is few and performance decrement does not happen are listed. As demerits, the points that it is impossible to made different kinds of virtual OS's move on the host OS and the obstacle of the host OS may have an influence on all virtual OS's are listed.

Next, there are there methods in server hardware virtualization as follows.

(a)     Virtualization by a firm ware

(b)     Virtualization by moving multiple gest OS's on the virtual software on a host OS

(c)     Virtualization by moving multiple gest OS's on the Hypervisor

In method (a), virtualization mechanisms are incorporated into the hardware of a server machine and hardware resources such as CPU and RAM are controlled by use of its mechanism. As examples of server machines within virtualization mechanisms, HP Integrity (Hewlett-Packard) [31] and System p (IBM) [32] are listed.

In method (b), hardware resources managed by a host OS on a server machine are divided in to virtual hardware resources and multiple gest OS's use them. As software to realize it, VMware Virtual Server (Microsoft) and Server (VMware) are listed.

As method (c), there are full virtualization and para virtualization. Both methods realize division use of hardware resources by operating multiple gest OS's on the Hypervisor which is the OS specialized in hardware control. Full virtualization means emulation of the hardware. The word of "emulation" means the realization of hardware resource's functions such as CPU and RAM in the state of software by incorporating device drivers in the Hypervisor. This method has the restriction that hardware resources such as a graphics board and a network interface must correspond to the Hypervisor. However, if this condition is met, it is possible to use any kinds of gest OS's on a virtual server without correction of the kernel, and each gest OS operates without being aware of being virtual environment. On the other hand, there is the fault that additional overhead processes generate in a CPU by using the Hypervisor. As examples of the software to realize this method, VMware ESXi (VMware) [33] and KVM (Qumranet) [34] are listed. Para virtualization prepares the OS for managing device drivers, and each guest OS accesses devices such as a network interface through it. Different from full virtualization, para virtualization needs to modify the kernels of the guest OS's when CPUs with a virtualization support function such as Intel-VT [35] and AMD-V [36] are not used. These technologies explained so far are used to create states same as what the multiple OS's operate virtually on a server machine. Essentially, these are technologies to gather multiple servers on one server machine.

As another virtual technology, service virtualization is listed. Network servers such as Web server, SMTP server, FTP Server have a virtualization function. For example, the virtualization function in a Web server makes it possible to display the different Web page on a Web Browser from the Web pages stored in each directory for each Web site. For example, when URL A (http://domain-a/) is inputted into the Web Browser, the contents of an html file (index.html) in directory-a are displayed on it. Next, when URL B (http://domain-b/) is inputted into the Web Browser, the contents of the html file (index.html) in directory-b are displayed on it. Thus, the function of managing multiple Web sites on one Web server is called the virtual host function.

However, when the Web servers are built by using virtualization technologies in this section, it is possible only to display the Web pages stored in one directory by using one URL. Even if theses Web pages are replaced to the FUCGIP, it is the same thing. Only one FUCGIP can be referred by using one URL. By using the virtual use method of CGI shown in this paper, when multiple users input same URL to the Web Browser, each user can access the FUCGIP stored in each directory prepared for each user. At the same time, the effect that is equal to add the user certification and access control is provided.

## 3. CONTENTS OF DACS SCHEME

### 3.1. Existing PBNM

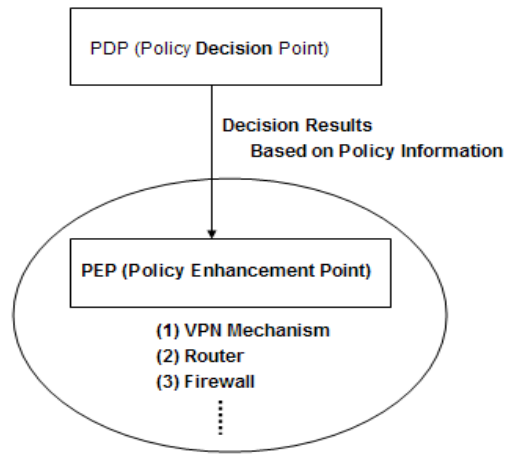

Figure 2 PBNM in IETF

As the work of network management, there are various kind of works such as the server load distribution technology [1][2][3], VPN (Virtual Private Network)[4][5] and network protocols[6][7]. However, these works are performed forward the specified different goal, and don't have the purpose of effective whole network management. As the work for managing a whole network, there is the work of Opengate[8][9], which controls Web accesses from LAN (Local Area Network) to internet. This work has the limited purpose of controlling Web access to internet. As the work for managing a whole network effectively without the limited purpose, there is the work of a PBNM (Policy-based network management) The PBNM's standardization is performed in various organizations. In IETF, a framework of PBNM [10] was established as shown in Figure 2. Standards about each element constituting this framework are as follows. As a model of control information stored in the server storing control information called Policy Repository, Policy Core Information model (PCIM) [11] was established. After it, PCMIe [12] was established by extending the PCIM. To describing them in the form of Lightweight Directory Access Protocol (LDAP), Policy Core LDAP Schema (PCLS) [13] was established. As a protocol to distribute the control information stored in Policy Repository or decision result from the PDP to the PEP, Common Open Policy Service (COPS) [14] was established. Based on the difference in distribution method, COPS usage for RSVP (COPS-RSVP) [15] and COPS usage for Provisioning (COPS-PR) [16] were established. RSVP is an abbreviation for Resource Reservation Protocol. The COPS-RSVP is the method as follows. After the PEP having detected the communication from a user or a client application, the PDP makes a judgmental decision for it. The decision is sent and applied to the PEP, and the PEP adds the control to it. The COPS-PR is the method of distributing the control information or decision result to the PEP before accepting the communication.

Next, in DMTF, a framework of PBNM called Directory-enabled Network (DEN) was established. Like the IETF framework, control information is stored in the server storing control information called Policy Server which is built by using the directory service such as LDAP [18], and is distributed to network servers and networking equipment such as switch and router. As the result, the whole LAN is managed. The model of control information used in DEN is called Common Information Model (CIM), the schema of the CIM（CIM Schema Version 2.30.0）[19] was opened. The CIM was extended to support the DEN [17], and was incorporated in the framework of DEN.

In addition, Resource and Admission Control Subsystem (RACS) [20] was established in Telecoms and Internet converged Services and protocols for Advanced Network (TISPAN) of European Telecommunications Standards Institute (ETSI), and Resource and Admission Control Functions (RACF) [21] was established in International Telecommunication Union Telecommunication Standardization Sector (ITU-T).

However, the PBNM has two structural problems such as communication concentration from many clients to a communication control mechanism called Policy Enforcement Point (PEP) and the necessity of the network updating at the time of introducing the PBNM into LAN. Moreover, it is often difficult for the PBNM to improve the user support problems in campus-like computer networks explained above. To solve these problems, we have studied about the next generation policy-based network management called the DACS Scheme.

### 3.2. Basic Principle of DACS Scheme

Figure 3 shows the basic principle of the network services by the DACS Scheme. At the timing of the (a) or (b) as shown in the following, the DACS rules (rules defined by the user unit) are distributed from the DACS Server to the DACS Client.
(a) At the time of a user logging in the client.

(b) At the time of a delivery indication from the system administrator.

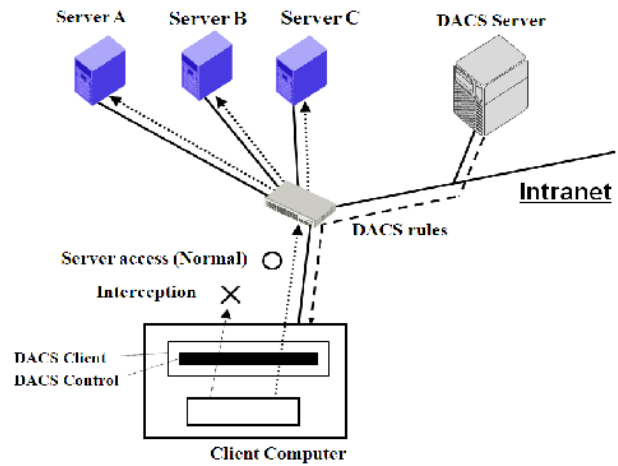

Figure 3  Basic Principle of the DACS Scheme

According to the distributed DACS rules, the DACS Client performs (1) or (2) operation as shown in the following. Then, communication control of the client is performed for every login user.

(1) Destination information on IP Packet, which is sent from application program, is changed.

(2) IP Packet from the client, which is sent from the application program to the outside of the client, is blocked.

An example of the case (1) is shown in Figure 3. In Figure 3, the system administrator can distribute a communication of the login user to the specified server among servers A, B or C. Moreover, the case (2) is described. For example, when the system administrator wants to forbid a user to use MUA (Mail User Agent), it will be performed by blocking IP Packet with the specific destination information.

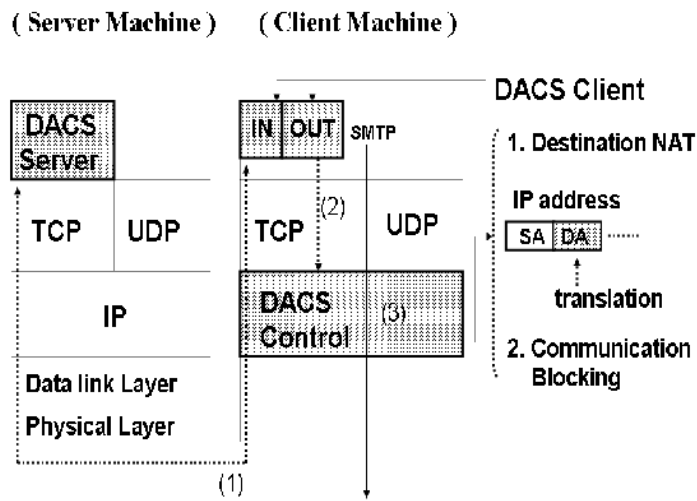

Figure4  Layer Setting of the DACS Scheme

In order to realize the DACS Scheme, the operation is done by a DACS Protocol as shown in Figure 4. As shown by (1) in Figure 4, the distribution of the DACS rules is performed on communication between the DACS Server and the DACS Client, which is arranged at the application layer. The application of the DACS rules to the DACS Control is shown by (2) in Figure 4. The steady communication control, such as a modification of the destination information or the communication blocking is performed at the network layer as shown by (3) in Figure 4.

### 3.3. Communication Control on Client

The communication control on every user was given. However, it may be better to perform communication control on every client instead of every user. For example, it is the case where many and unspecified users use a computer room, which is controlled. In this section, the method of communication control on every client is described, and the coexistence method with the communication control on every user is considered. When a user logs in to a client, the IP address of the client is transmitted to the DACS Server from the DACS Client. Then, if the DACS rules corresponding to IP address, is registered into the DACS Server side, it is transmitted to the DACS Client. Then, communication control for every client can be realized by applying to the DACS Control. In this case, it is a premise that a client uses a fixed IP address. However, when using DHCP service, it is possible to carry out the same control to all the clients linked to the whole network or its subnetwork for example. When using communication control on every user and every client, communication control may conflict. In that case, a priority needs to be given. The judgment is performed in the DACS Server side as shown in Figure 5. Although not necessarily stipulated, the network policy or security policy exists in the organization such as a university (1). The priority is decided according to the policy (2). In (a), priority is given for the user's rule to control communication by the user unit. In (b), priority is given for the client's rule to control communication by the client unit. In (c), the user's rule is the same as the client's rule. As the result of comparing the conflict rules, one rule is determined respectively. Those rules and other rules not overlapping are gathered, and the DACS rules are created (3). The DACS rules are transmitted to the DACS Client. In the DACS Client side, the DACS rules are applied to the DACS Control. The difference between the user's rule and the client's rule is not distinguished.

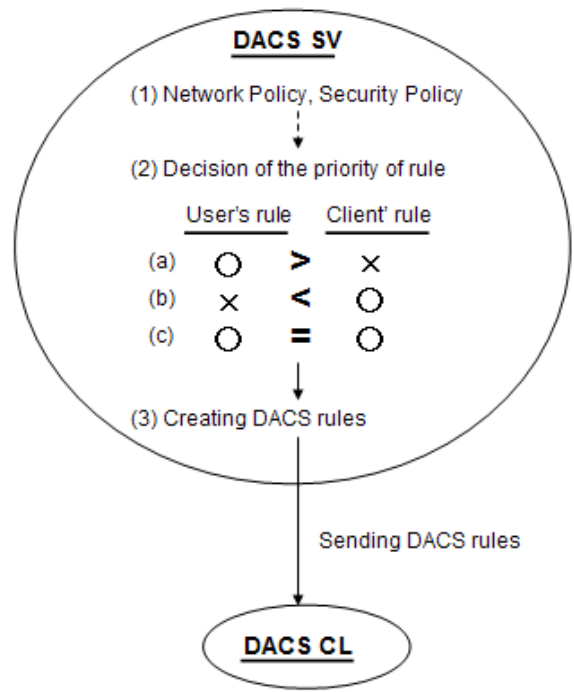

Figure 5 Creating the DACS rules in the DACS Server side

## 3.4. Security Mechanism of the DACS Scheme

In this section, the security function of the DACS Scheme is described. The communication is tunneled and encrypted by use of SSH. By using the function of port forwarding of SSH, it is realized to tunnel and encrypt the communication between the network server and the, which DACS Client is installed in. Normally, to communicate from a client application to a network server by using the function of port forwarding of SSH, local host (127.0.0.1) needs to be indicated on that client application as a communicating server. The transparent use of a client, which is a characteristic of the DACS Scheme, is failed. The transparent use of a client means that a client can be used continuously without changing setups when the network system is updated. The function that doesn't fail the transparent use of a client is needed.

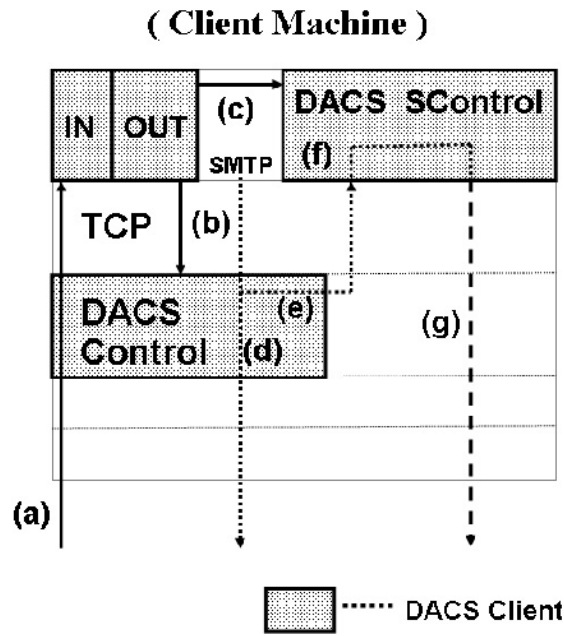

Figure 6 Extend Security Function

The mechanism of that function is shown in Figure 6. The changed point on network server side is shown as follows in comparison with the existing DACS Scheme. SSH Server is located and activated, and communication except SSH is blocked. In Figure 6, the DACS rules are sent from the DACS Server to the DACS Client (a). By the DACS Client that accepts the DACS rules, the DACS rules are applied to the DACS Control in the DACS Client (b). The movement to here is same as the existing DACS Scheme. After functional extension, as shown in (c) of Figure 6, the DACS rules are applied to the DACS SControl. Communication control is performed in the DACS SControl with the function of SSH. By adding the extended function, selecting the tunneled and encrypted or not tunneled and encrypted communication is done for each network service. When communication is not tunneled and encrypted, communication control is performed by the DACS Control as shown in (d) of Figure 6. When communication is tunneled and encrypted, destination of the communication is changed by the DACS Control to localhost as shown in (e) of Figure 6. After that, by the DACS STCL, the communicating server is changed to the network server and tunneled and encrypted communication is sent as shown in (g) of Figure 6, which are realized by the function of port forwarding of SSH. In the DACS rules applied to the DACS Control, localhost is indicated as the destination of communication. In the DACS rules applied to the DACS SControl, the network server is indicated as the destination of communication. As the functional extension explained in the above, the function of tunneling and encrypting communication is realized in the state of being suitable for the DACS Scheme, that is, with the transparent use of a client. Then, by changing the content of the DACS rules applied to the DACS Control and the DACS SControl, it is realized to distinguish the control in the case of tunneling and encrypting or not tunneling and encrypting by a user unit. By tunneling and encrypting the communication for one network service from all users, and blocking the untunneled and decrypted communication for that network service, the function of preventing the communication for one network service from the client, which DACS Client is not installed in is realized. Moreover, even if the communication to the network server from the client, which DACS Client is not installed in is permitted, each user can select whether the communication is tunneled and encrypted or not. The function of preventing information interception is realized.

## 4. SYNOPSIS OF THE DACS WEB SERVICE

### 4.1. Two Kinds of Functions of Web Service Based on DACS Scheme

Two kinds of functions of Web Services based on DACS Scheme are described, here.

At First, the function to use data from database is developed. To realize this function, DACS Scheme needs to be extended, and the program on Web Server needs to be implemented in correspondence to the extended DACS Scheme as shown in Figure 7. In the network with DACS Scheme, after a user's logging in a client (a), user name and IP address are sent to DACS SV (b). Then, DACS rules are sent back to DACS CL (c). Moreover, user name and IP address are sent to the program on Web Server. Then, the server side program on Web Server can identify the user by checking the login information and the source IP address from the client, and can change the processing of the program every user. When each different user accesses the program with same URL, different information for each user can be searched and extracted from database, and be displayed on Web Browser. Through the processing from (1) to (7), this new function is performed.

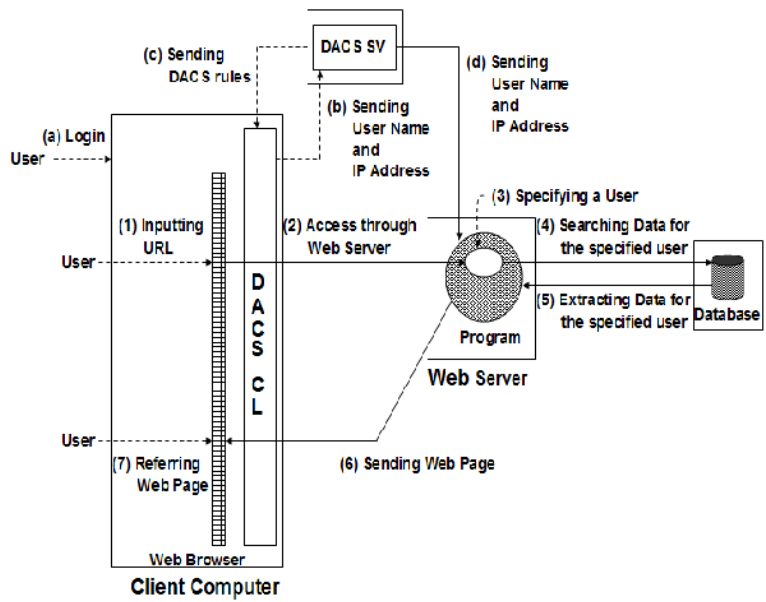

Figure 7 Function Using Data from Database

Next, the function to use data from document medium is developed for the respective user. In the network with DACS scheme, different IP address and TCP port can be assigned for one host name by a user unit. Therefore, different document medium with same file name on different Web Server can be referred for each user by inputting same URL to Web Browser. When this principle is combined with the function of virtual host that is equipped as Web Server, it is possible to use Web Server as shown in Figure8.

By the function of virtual host, multiple groups of socket (IP address and TCP port) can be assigned for one Web Server. The referred document can be changed every socket. First, in Document root of Web Server in Figure 9, directories (Dir A,B,C,D….) are prepared for each user. By the function of virtual host, each directory is connected to each socket as one pare. By changing TCP port number (3000,3001,3002….) for one IP address (192.168.1.1), sockets

International Journal of Computer Networks & Communications (IJCNC) Vol.4, No.1, January 2012

corresponding to each directory are prepared. Next, movement on this mechanism is described. One user inputs one URL to Web Browser. When the URL is inputted by User A, the file in Dir A that is connected to the socket (192.168.1.1:3000) is referred. Equally, when by User B, the file in Dir B that is connected to the socket (192.168.1.1:3001) is referred. When by User C, the file in Dir C that is connected to the socket (192.168.1.1:3002) is referred. When the document medium with same name exists in each directory (Dir A,B,C….), each user can see different contents by inputting same URL to Web Browser. For information sender, because it is possible to notify information to the specific user by uploading document medium to the predetermined directory, information usage becomes largely wide. Because information sender can describe the content of document medium easily and freely, it is possible to communicate the information with much expressive power and impact.

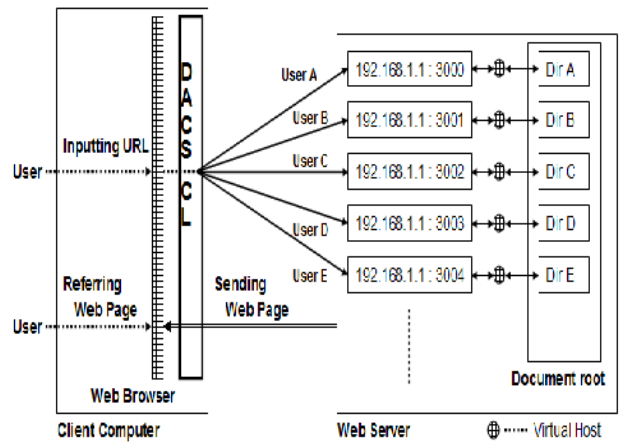

Figure 8 Function Using Data from Document Medium

As the result, by letting both functions coexist as shown in Figure 9, the Web Service that a user can use information on the network regardless of information storage form is realized.

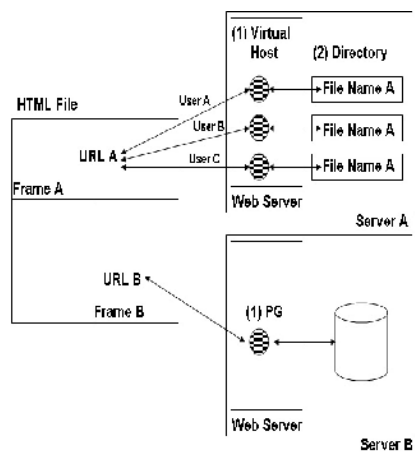

Figure 9 Web Service by two functions

### 4.2. Contents of the DACS Web Service

In Figure 10, synopsis of DACS Web Service is shown. The function to use data from database of information system such as a system managing results for a student, is shown as Function α. The function to use data from document medium such as a simple text file and a PDF file, is shown as Function β. After a user's inputting URL into Web Browser, communication control by DACS CL (DACS CTL) is performed. As the result, function α or Function β is used. Because the function of either is automatically selected every each URL according to DACS rules, a user can use data from information system or document medium dispersing on the network without being conscious of that function is used. In other words, a user can use information regardless of storage form and storage place of data freely and easily, if a user knows URL and the kind of information acquired by that URL. Even if whichever of Function α or Function β is used, data is displayed on Web Browser after inputting URL. Three kinds of data, which are sent by a user unit (a) and by a group unit (b) and by all users unit, are displayed.

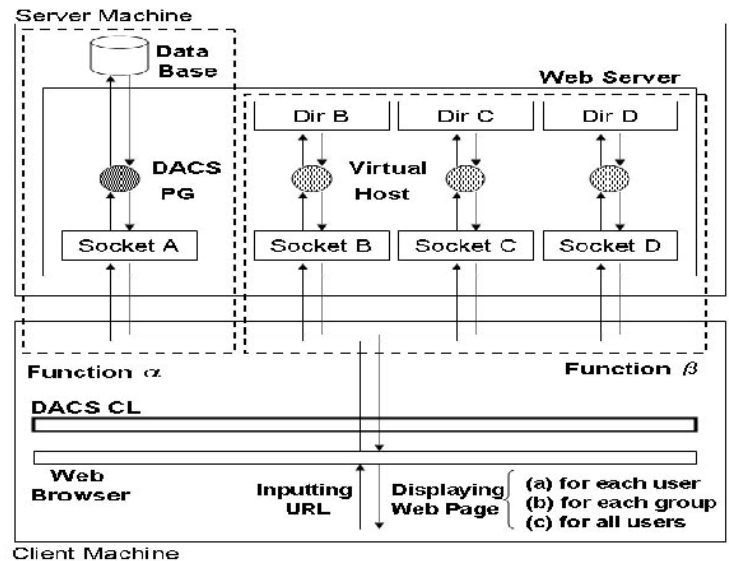

Figure 10 DACS Web Service

Here, details of Function α are shown. After extension, the functions of retrieving data for each group (Function α2) or for all users (Function α3) can be used. There are differences among Function α1, Function α2, and Function α3 in the program extracting data from a database for a request from a Web browser. In the program of Function α1, data is extracted for each user, as shown by (1). In the program of Function α2, data is extracted for each group, as shown by (2). In the program of Function α3, data is extracted for all users, as shown by (3). In the existing function to retrieve data from a database, as shown by (1), it is possible to specify which user is sending communication through the Web browser. Therefore, the function is extended to set a correspondence list of the user and group name in the DACS Server and send that correspondence list from the DACS Server to the program of Function α2. As a result, because the program of Function α2 can recognize the group to which a user belongs, it is possible to extract information for each group. Even if a user belongs to multiple groups, it is possible to extract the data of all groups. In addition, it is possible to extract the data of a specific group by sending its group name as a parameter of the URL. In the program of Function α3, data is extracted for all of these users.

Because it is the function of a normal Web Service that does not introduce DACS Scheme, it is generally realized without a technical problem.

Next, details of Function are shown. Function 1 displays data of the document medium dynamically for each user. By use of this function, the function for each group (Function 2) and the function for all users (Function 3) are realized. Function 2 relates the URL for each group (Group URL1, Group URL2….) to each document, which is stored in each directory for each group. Function 3 relates the URL for all users (All Users URL) to the document, which is stored in the directory for all users. To send information, only uploading a file as a document medium into the predetermined directory (directory for each user, directory for each group, and directory for all users) is necessary. Information for each group can be recieved by the specific URL for each group. In addition, the users not belonging to each group can not access it by using the URL. Information for all users can be recieved by use of the URL for all users. By using the DACS Web Service, not only information for each user but also information for each group and for all users, can be used from the document medium.

## 5. VIRTUAL USE METHOD OF THE FUCGIP

### 5.1. Principle of Virtual Use method of the FUCGIP

In this paper, the method that is realized by the Function 1 is proposed. By using this function, FUCGIP I is accessed virtually through same URL from users in each group. To be concrete, this method is realized by the following procedure.

**(Step1) First setting of the FUCGIP**

First, the FUCGIP is set by a normal procedure. For example, the program files as the FUCGIP are placed on the Web Server, and initial setting is performed. For example, the setting of initial parameter of the FUCGIP and permission of the program files. As the result, users can use the programs of the FUCGIP by inputting one URL into a Web Browser.

**(Setp2) Setting for virtual use of the FUCGIP programs**

After copying the directory that stores the programs, it is pasted as another directory with another name. By repeating a similar operation, multiple directories for each group are prepared. At the same time, the content of the DACS rules is changed. As the result, users that belong to same group become possible to access the programs of same directory by use of a URL. On the other hand, users that belong to other group become impossible to access the above programs by use of same URL. By these procedures, in the form of using same URL, users in each group can access the programs of the directory in each group, and cannot access the programs of other group. Virtual use of the FUCGIP is realized without a special mechanism.

A concrete example of it is shown in Figure 11. As first step, the program X as the FUCGIP and other files such as data file are placed in directory A (Dir A in Figure 11), and initial setting of it is performed. As the result, users can access and use it. Next, second step is as follows. At first, Dir A is copied and pasted with another name. In Figure 11, Dir B and Dir C are the pasted directories. Each directory is named with the regularity. Though each socket is connected to each directory through the virtual host by the system setting, each name is allocated to be easy to automate the setting. At the same time, by changing the DACS rules, the host name in URL and the communication port is converted to each socket every group. In Figure 11, when users in Group A inputs one URL into a Web Browser, they access the program X in Dir A through by way of 192.168.1.1:3000. In the case of users in Group B, they access the program X in Dir B through by way of 192.168.1.1:3001. In the case of users in Group C, they access the program X in Dir C by way of 192.168.1.1:3002. Then, users in Group A con not access the program in Dir B and Dir C. Users in Group B con not access the program in Dir A and Dir C. Users in Group C con not access the program in Dir A and Dir B. In this way, virtual use of the CGI program is realized simply.

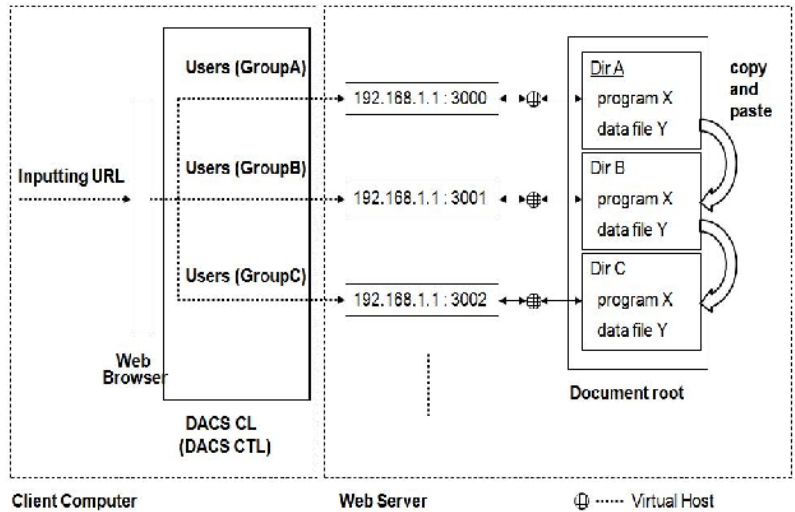

Figure 11 Virtual Usage of the CGI program

## 5.2. Functional Experiments

To confirm the possibility of the proposed method, a functional experiment was done. The experiment system is described in figure 11. In this figure, there are three system components, which are the Web Server to set the FUCGIP and the DACS Server and the client computer with the DACS Client. As an example of the FUCGIP, access counter program was set on the Web Server such as Figure 11. First, one user belonging to GroupA logged in the client computer and inputted the URL for accessing the access counter into the Web Browser. Then, the number of 11 was displayed on the Web Browser. Next, another user belonging to GroupB logged in the client computer and inputted the URL for accessing the access counter into the Web Browser. As the result, the number of 5 was displayed on the Web Browser. Because the small number was displayed in comparison with a former user, they accessed the different FUCGIP clearly stored in the different directory. That is, when the different users accessed the access counter by use of same URL, they accessed another FUCGIP stored in different directory.

## 6. CONCLUSION

In this paper, virtual use method in common gateway interface program on the DACS Scheme was shown. This is realized in the case of introducing the DACS Scheme as a network management scheme, which is a kind of PBNM that we have been proposed before. The DACS Scheme is the scheme which has a control mechanism on each client as the PEP for communications. By using the DACS Scheme, the FUCGIP is used virtually and securely without changing the physical constitution of the network. To be concrete, when multiple users input same URL to the Web Browser, each user can access the FUCGIP stored in each directory prepared for each user. At the same time, the effect that is equal to add the user certification and access control is provided. For the points, we prepare experiment environment, and perform simple functional experiments to confirm the possibility of this method. In the future study, experiments by using various kinds of CGI programs will be performed. That is, studies to raise the effectiveness of this method will be performed.